\newcommand{\da}{\partial}

\newcommand{\un}{\underline}

\newcommand{\hf}{{_1\over^2}}
\newcommand{\ov}{\overline}

\newcommand{\qq}{\begin{equation}}
\newcommand{\qqq}{\end{equation}}
\newcommand{\myeqnarray}[1]{
  \begingroup
  \jot=#1pt
  \arraycolsep=2pt
  \begin{eqnarray}}
\newcommand{\beqnarray}{\myeqnarray{3}}
\newcommand{\eeqnarray}{\end{eqnarray}\endgroup}
\newcommand{\ba}{\begin{array}{cc}}
\newcommand{\ea}{\end{array}}

\documentclass[12pt]{article}
\usepackage[xdvi]{graphicx}%
\usepackage{dcolumn}
\usepackage{epsfig}
\usepackage{feynmf}
\begin{document}
\title{On the local space-time structure of non-equilibrium steady states.}
\author{Rapha\"el Lefevere\footnote{lefevere@math.jussieu.fr}}
\date{}
\begin{center}
{\large{\bf On the local space-time structure \\ of non-equilibrium steady states.}}\\
\vspace{5mm}
Rapha\"el Lefevere\footnote{lefevere@math.jussieu.fr}\\
Laboratoire de Probabilit\'es et mod\`eles al\'eatoires\\
UFR de Math\'ematiques Universit\'e Paris 7, 
\\ Case 7012, 75251 Paris Cedex 05, France  
\end{center}

\begin{abstract}
\noindent We consider the Gibbs representation over space-time of non-equilibrium dynamics of Hamiltonian systems defined on a lattice with local interactions.   We first write the corresponding action functional as a sum of local terms, defining a local action functional.  We replace the local system by a translation-invariant system whose dynamics has  an identical space-time characterization. We study in details the irreversible properties of the new dynamics, define the local conductivity and show its equivalence with the Green-Kubo formula.
Given the definition of the local heat conductivity and using conservation of energy, we derive the shape of the temperature profile. Next, we find an explicit formula for the non-equilibrium stationary measure of harmonic systems.
Finally, we apply our scheme to various approximations of anharmonic Hamiltonian models, show how to compute their thermal conductivity and recover results confirmed in numerical simulations.
\end{abstract}

\newpage

\section{Introduction.}
Non-equilibrium stationary states (NESS) of systems  of classical Hamiltonian oscillators located on a lattice coupled to heat baths at their boundaries have been extensively studied over recent years \cite{eck1,eck3,ReybelletThomas0}.  Numerous global results concerning those NESS have been obtained, including existence and uniqueness of the stationary probability measure. The positivity of the entropy production and the validity of the Gallavotti-Cohen fluctuation theorem has also been established \cite{Eckent,Maes2, ReybelletThomas1}.  They have also been abundantly studied with the help of numerical simulations, see \cite{bonetto,livi} for a review of those and a general overview of the subject.

However, in contrast to equilibrium states for which the explicit Gibbs formula may be used, the effective computation of correlation functions of the stationary states out of equilibrium remains a challenging problem.  Conceptually, one may distinguish two different reasons for that.  The first one is common with the equilibrium situation and has to do with the difficulty of dealing with nonlinear interactions between the components of the system.  The second one has to do with the lack of translation invariance which is, in a sense built-in in the non-equilibrium set-up.  This paper adresses the second issue for Hamiltonian systems defined on a lattice with local interactions.
The difficulties we have just mentioned  are particularly obvious when one tries to identify the physical mechanism giving rise to a finite thermal conductivity in  non-linear Hamiltonian systems and compute its dependance on the microscopic interactions.  The conductivity is a local property of the system that relates local quantities like the current and the local temperature gradient.  The goal of this paper is to outline a method which, by dealing with the lack of translation invariance of the system, allow to focus on the effect of nonlinearities on the non-equilibrium steady states and perform concrete computations of their correlation functions.  The method is a priori independent of any approximation scheme and apply to any local Hamiltonian lattice system out of equilibrium. 

As we will argue later, what matters locally and away from the boundaries in a non-equilibrium set-up is not so much the spatial geometry but rather the specific way in which the time reversal symmetry of the equilibrium  dynamics is broken. It turns out that heuristically, the non-equilibrium dynamics may be characterized locally in a simple, universal and translation-invariant way, see (\ref{reversalbasic}) below.  This follows from local (on the lattice) conservation of energy and the characterization of NESS in terms of probabilities over pathspace introduced by Maes \cite{Maes0}. Therefore, in order to study the local dynamics, we will consider a spatially homogeneous Hamiltonian chain of oscillators coupled to a stochastic thermostat at a fixed temperature $T$ and,  in order to recover the time-reversal symmetry breaking induced locally by the heat baths located at the boundaries, we include an additional non-Hamiltonian term in the deterministic part of the dynamics.   The construction of the dynamics is very similar in spirit to the  Evans heat flow algorithm \cite{Evans} but ours proceeds through the introduction of a {\it stochastic} thermostat, which makes analytical computation straightforward.  In particular, the proof that the dynamics satisfy the Gallavotti-Cohen fluctuation theorem is simple and the equations for the correlation functions in the stationary state may be easily written down. Systems with  Langevin stochastic thermostats  and non-conservative forces producing entropy have been considered in  \cite{Kurchan,Maes1}.  Also, when comparing our dynamics with the Nos\'e-Hoover type thermostats (i.e. purely deterministic), it is important to notice that the additional deterministic term is itself time-reversible.  It is only when coupled to the Langevin thermostat that the dynamics becomes irreversible in time, in a stochastic sense. We shall see how to recover the  linear response theory's Green-Kubo formula in a natural way. However, our scheme hopefully also allows to describe the behaviour of systems far from equilibrium.  We shall also see how to reconstruct the temperature profile on the original chain coupled to heat baths at different temperatures.  
The basic assumption that is made is that two systems having the same time-reversal symmetry breaking functional will have basically the same average heat current.  The validity of this assumption would certainly benefit from more of theoretical and experimental arguments.

In the last section, we consider a concrete anharmonic Hamiltonian model, generic of models having a normal heat conduction behaviour.  The Hamiltonian is given by,
\qq
H(\un q,\un p)=\sum_{i=1}^N \big[\frac{p_i^2}{2}+\omega^2\mu^2\frac{q_i^2}{2}+\frac{\lambda}{4}q_i^4+\frac{\omega^2}{2}(q_i-q_{i-1})^2\big]
\label{basicmodel}
\qqq
On that example, we will see how to implement our strategy and study the problem with the help of the non-equilibrium translation invariant dynamics we have defined before.  We will show that  this dynamics provides a source term, absent in equilibrium, in the equation for the evolution of the average heat current $j$ in the chain.   This is of course what should be expected because the dynamics is built so as to break the time-reversal symmetry of the equilibrium dynamics in a proper way. 
Roughly speaking we get an equation of the form,
\qq
\frac{d}{dt}j(t)=-\gamma j(t)+\lambda {\cal N}(t) + S(t).
\label{rough}
\qqq
The first term on the RHS of this equation is a damping term coming
from the coupling of the chain to an external  friction represented by
the coefficient  $\gamma$.  The second term comes from the Hamiltonian
evolution and vanishes when the anharmonic part of the interaction is
zero, i.e., when $\lambda=0$.  The third term represents an external
``creation" of current and comes from the action of the non-equilibrium dynamics. Equation
(\ref{rough}) will therefore yield a balance equation for the heat current in the stationary state. 
First, we review the linear case ($\lambda=0$), for which we give an explicit formula for the stationary measure out of equilibrium.
Next we study two approximations which have been used recently to analyze the effect of the anharmonic terms on the non-equilibrium properties of the Hamiltonian system.
The first one \cite{Bernardin, Kipnis} amounts to replace the effects of the anharmonic terms by a random exchange of energy between nearest-neighbours, this gives rise to a normal heat conductivity and to a linear profile of temperature.  

The second one has been derived and used over the years since Peierls \cite{Bricmont,Horie, Kwok, LefevereSchenkel,Peierls,Spohn}. It amounts to a closure assumption on the hierarchy of correlations and to a derivation of a stationary Boltzmann-type equation with a  collision kernel representing the interaction between the phonons due to the nonlinearity of the system.  The basic mechanism which gives rise to a normal heat conductivity has been known to physicists for a long time, it was postulated to be the result of some particular resonances in the collision between phonons.  However, the lack of explicit localization and of a proof of the existence of those resonances  prevented the effective computation of the thermal conductivity in terms of the physical microscopic parameters. The localization of the resonances in the model (\ref{basicmodel}) for large pinning $\mu$ has been achieved in \cite{LefevereSchenkel}, where a similar translation invariant dynamics has been devised, and the value
\qq
\kappa=\alpha\frac{\omega^9\mu^3}{\lambda^2T^2}
\label{conductivity}
\qqq
for the thermal conductivity was obtained.  The value of $\alpha=0.275637$ was computed in \cite{AokiSpohn} where the expression  (\ref{conductivity}) was also tested and numerically confirmed.  The localization of the resonances at $\mu=0$ was obtained by Pereverzev \cite{Pereverzev} who used it to discuss anomalous heat conduction in a related anharmonic model.
 
\section{Local time-reversal in  models for heat conduction.}

In order to explain the behaviour of the thermal conductivity in crystalline solids,
one often modelizes the solid by a chain (or lattice in higher dimension) whose ends
are coupled to heat baths maintained at different temperatures.  The coupling can be
taken stochastic and more precisely of Langevin type. In one dimension, the set-up is
as follows. At each site $i$ of a lattice $\{1,\ldots,N\}$ is attached a particle
of momentum $p_i$ and position $q_i$. The dynamics is hamiltonian in the bulk and
stochastic through the Langevin coupling to heat baths at the boundaries. The
Hamiltonian is of the form,
\qq
H(\un p,\un q)=\sum_{i=1}^{N}\Bigl(\hf p_i^2 +V(q_i)\Bigr)+\sum_{i=2}^N
U(q_i-q_{i-1})+U(q_1)+U(q_N).
\label{Ham}
\qqq
The equations
of motions are given by,
\beqnarray
dq_i&=&p_idt,\quad i=1,\dots,N,
\nonumber
\\
dp_i&=&-\frac{\da H}{\da q_i}(\un p,\un q)dt,\quad i=2, \ldots,N-1,
\label{dynamics0}
\eeqnarray
and,
\beqnarray
dp_{1}&=&-\frac{\da H}{\da q_{1}}(\un p,\un q)dt-\gamma p_{1} dt
+\sqrt{2\gamma kT_L}\,dw_{l}\,,
\nonumber
\\
dp_{N}&=&-\frac{\da H}{\da q_N}(\un p,\un q)dt-\gamma p_{N} dt
+\sqrt{2\gamma kT_R}\,dw_{r}\,.
\label{dynamics00}
\eeqnarray
$T_L$  and $T_R$ stand for the temperature of the left and
right reservoirs, respectively,
whereas  $w_{l}$ and $w_r$ are two independent standard Wiener processes.

It is an easy fact to check that when $T_L=T_R=T=\beta^{-1}$,
the equilibrium measure on the configuration space ${\bf R}^{2N}$ whose density
with respect to the Lebesgue measure is given by
\qq
\rho(\un p,\un q)=Z^{-1}e^{-\beta H(\un p,\un q)}
\qqq
is invariant (stationary) for the stochastic dynamics defined above.
In the case of two different temperatures, existence, uniqueness and
exponential convergence to an unique invariant state has been established under fairly
general conditions on the potentials $U$ and $V$ \cite{eck1,eck3, ReybelletThomas0}. In the
case of harmonic coupling, the covariance of the stationary state has been exactly
computed in \cite{nak,rll}.

In order to describe the conduction of heat in the crystal, one defines a local energy function,
\qq
h_i(\un p,\un q)=\frac{p_i^2}{2}+V(q_i)+\hf(U(q_{i+1}-q_i)+U(q_i-q_{i-1}))
\label{local}
\qqq
for $i\neq 1,N$,
\qq
h_1(\un p, \un q)=\frac{p_1^2}{2}+V(q_1)+U(q_1)+\hf U(q_{2}-q_1)
\qqq
and similarily for $h_N$.
The heat current is defined through the time evolution of the local energy,
\qq
\frac{d h_i}{dt}=j^+_i-j^{-}_{i}
\label{conservation0}
\qqq
for  $i\neq 1,N$ and where,
\qq
j^+_i=\hf F(q_{i}-q_{i+1})(p_i+p_{i+1}),
\label{current0}
\qqq
\qq
j^-_i=\hf F(q_{i-1}-q_{i})(p_i+p_{i-1}),
\label{current00}
\qqq
where $F=-U'$ and $j_i$ is defined to be the microscopic current of energy or heat between atom $i$ and $i+1$, i.e the rate of energy transfer between those atoms.  We observe that $j^+_{i-1}=j^-_{i}$ and in the following we shall use the shorthand notation, $j_i\equiv j^+_i$.

\noindent We use the notation $\left<.\right>$ to denote average with respect to the stationary state.
And define the local temperature in the chain to be
\qq
\left<p_i^2\right>\equiv T_i
\label{temperature}
\qqq
and the average heat current is,
\qq
j\equiv\left<j_i\right>=\frac{\omega^2}{2}\left<F(q_i-q_{i+1})(p_i+p_{i+1})\right>
\qqq
which must be constant throughout the chain due to conservation of energy.  

\noindent This follows from,
\qq
\left<\frac{d h_i}{dt}\right>=\left<j^+_i\right>-\left<j^{-}_{i}\right>=0
\label{conservation000}
\qqq
and $j^+_{i-1}=j^-_{i}$.

\noindent Fourier law states that,
\qq
j=\kappa(T_i)(T_{i+1}-T_i)
\qqq
where $\kappa$ is the conductivity of the crystal, a major problem is to understand how the anharmonic part of the Hamiltonian makes this constant finite and to compute its value from the microscopic parameters.

An elegant and systematic  way of describing non-equilibrium dynamics was devised by Maes \cite{Maes0}.  For stochastic processes, it amounts to compare the probability of trajectories of the non-equilibrium process at hand and of its time-reverse. By the Girsanov formula, the ratio of the two probabilities is given by the exponential of an ``action functional".  Then the part of the action functional which breaks the time-reversal invariance is identified with the entropy production. Let $P_{\mu}^t$ the pathspace measure on trajectories $\xi=((\un q(s),\un p(s)), s\in [-t,t])$ obtained from the dynamics (\ref{dynamics0},\ref{dynamics00}) started from initial conditions which are distributed according to some measure $\mu$.  We define the time-reversal operator on trajectories as follows, $(\Pi\xi)_s=(\un q(-s),-\un p(-s))$ and let 
$P_{\mu}^t\Pi$ be the pathspace distribution of the time-reverse process.  In \cite{Maes2, ReybelletThomas1}, it has been show that the ratio of the two measures may be written as,
\qq
\frac{dP_{\mu}^t}{dP_{\mu}^t\Pi}=\exp(R^t(\xi))
\label{actionratio}
\qqq
modulo temporal boundary terms involving the distribution $\mu$.

\noindent The action functional $R^t$ is given by,
\beqnarray
R^t(\xi)&=&\beta_L(\int_{-t}^t j_1(s) ds-(h_1(\xi_t)-h_1(\xi_{-t}))
\nonumber
\\
&+&\beta_R(\int_{-t}^t- j_{N-1}(s) ds-(h_N(\xi_t)-h_N(\xi_{-t})),
\label{actiondifference}
\eeqnarray
with $\beta_L=T_L^{-1}$ and $\beta_R=T_R^{-1}$.
We observe now that for any collection $K_i$, $i=1,\ldots,N$, such that $K_1=T_L$ and $K_N=T_R$,
\qq
R^t(\xi)=\sum_{i=1}^{N-1}\int_{-t}^t \sigma_i(s)ds-\sum_{i=1}^{N}\frac{1}{K_i}(h_i(\xi_t)-h_i(\xi_{-t})),
\label{loceq}
\qqq
where we defined the local entropy production $\sigma_i$,
\qq
\sigma_i=(\frac{1}{K_i}-\frac{1}{K_{i+1}})j_i.
\qqq
This follows from {\it local} energy conservation inside the chain written under the integral form,
\qq
h_k(\xi_t)-h_k(\xi_{-t})=\int_{-t}^{t}(j_k(s)-j_{k-1}(s))ds
\qqq
Except for the arbitrariness of the $K_i$'s, this shows that the conservation energy imposes strict conditions on the representation of the action (\ref{actiondifference}) in terms of the local variables $j_i$ and $h_i$.

Among all the equivalent representations for the action difference (\ref{loceq}), we now {\it choose} one which is useful because it allows detailed local analysis of the dynamics.  Let us take $K_i=T(\frac{i}{N})$, with $T(.)$ a smooth function over the interval $[0,1]$.  Now, at lowest order in $N^{-1}$, (\ref{loceq}) becomes,
\qq
R^t(\xi)=\frac{1}{N}\sum_{i=1}^{N-1}\int_{-t}^t j_i(s)\frac{\nabla T(\frac{i}{N})}{T^2(\frac{i}{N})}ds-\sum_{i=1}^{N}\beta(\frac{i}{N})(h_i(\xi_t)-h_i(\xi_{-t})).
\qqq
In the limit $N\rightarrow\infty$ and around a fixed point $x$ of the interval $[0,1]$, the variation of the coefficient in front of the integral of the current and the local energy goes to zero.   So we subdivide the interval $[0,1]$ in subintervals of size $\epsilon$, with $\epsilon$ small but $N>>\epsilon^{-1}$, each interval being centred about some point that we denote by $ x_k$.
Now, at lowest order in $\epsilon$, we write,
\qq
R^t(\xi)=\sum_k R^t_{x_k}(\xi)
\label{loc1}
\qqq
The local action functional $R^t_{ x}(\xi)$ is given by,
\qq
R^t_{x}(\xi)=\frac{\nabla T(x)}{T^2(x)}\frac{1}{N}\sum_{i\in B_\epsilon(x)}\int_{-t}^t j_i(s)ds-\beta(x)\sum_{i\in B_\epsilon(x)}(h_i(\xi_t)-h_i(\xi_{-t}))
\label{loc2}
\qqq
where $B_\epsilon(x)=\{j| \,|j-Nx|\leq\hf N\epsilon\}$.  Note that $|B_\epsilon(x)|=\epsilon N$.
 So locally around the point $x$, we get the following characterization of the non-equilibrium dynamics.
\qq
Q_{x}=\exp\left(\frac{\nabla T(x)}{T^2(x)}\frac{1}{N}\sum_{i\in B_\epsilon(x)}\int_{-t}^t j_i(s)ds-\beta(x)\sum_{i\in B_\epsilon(x)}(h_i(\xi_t)-h_i(\xi_{-t}))\right).
\label{reversalbasic}
\qqq
with,
\qq
\frac{dP_{\mu}^t}{dP_{\mu}^t\Pi}=\prod_k Q_{x_k}
\qqq
The important feature of (\ref{loc2}) is that the original problem has become locally (in the box $B_\epsilon(x)$) translation invariant.
For a given $x$, we now fix $T\equiv T(x)$ and $\tau\equiv N^{-1}\nabla T(x)$, $M=[N\epsilon]$ and in the next section, we define a dynamics which has exactly the property (\ref{reversalbasic}) for those parameters, see (\ref{reversalbasic1}). Then, in the last section, we will write  the equations that the correlations of the stationary state satisfy in a concrete model.   This amounts to ask {\it local} stationarity in the original non-translation invariant chain and allows on one hand to interpret the parameter $T$ as a local temperature because it gives the average kinetic energy.  On the other hand, it fixes the current as a function of the temperature gradient $\nabla T$ and local temperature $T$.  Going back to the original problem, the fact that the average current in the stationary state is constant in position then fixes the kinetic energy (i.e. temperature) profile as we shall see at the end of section 5.  We have derived (\ref{reversalbasic}) for Hamiltonian containing only nearest-neighbour interactions.  Presumably the same identity may be derived for systems with sufficiently local interactions.  And in the next section, we devise a dynamics which include more general interactions.

\section{The dynamics and its entropy production.}
  
We consider a periodic lattice hamiltonian system described by the Hamiltonian,
\qq
H(\un q,\un p)=\sum_{i=1}^M \left[\frac{p_i^2}{2}+V(q_i)+\hf\sum_{k=1}^{M-1}(U^k(q_i-q_{i+k})+U^k(q_i-q_{i-k}))\right]
\label{Hamilton}
\qqq
with the convention $i+M=i$. Note that the form of the Hamiltonian is slightly more general than in the previous section but is the same when $U^k=0$, for $k\neq 1$.
Proceeding exactly as in (\ref{local},\ref{conservation0}), one computes the local energy variation rate, and by analogy,
\qq
j_i=\hf\sum_k(p_i+p_{i+k})F^k(q_{i}-q_{i+k}),
\label{current1}
\qqq
where $F^k=-(U^k)'$ .  $j_i$ is interpreted as the total heat current entering the site $i$ coming from the ``right" of the chain.
We shall use below the spatial average of the energy current,
\qq
J=\frac{1}{M}\sum_{i=1}^M j_i
\label{spatial}
\qqq
Our dynamics is,
\beqnarray
dq_i&=&p_i dt\nonumber
\\
dp_i&=&-\gamma p_i dt-\frac{\partial H}{\partial q_i} dt+\frac{\tau}{2T}\sum_k(F^k(q_{i-k}-q_i)+F^k(q_i-q_{i+k}))dt+\sqrt{2\gamma T}dw_i\nonumber\\
\label{dynamics}
\eeqnarray
and the $w_i$ are standard independent Brownian motion $i=1,\ldots,M$.    The term proportional to $\tau$ is the non-equilibrium part of the dynamics and is responsible for the breaking of the time-reversal symmetry of the equilibrium dynamics (at $\tau=0$).  We will see that its particular form allows to show the validity of (\ref{reversalbasic}) and thus the validity of the fluctuation theorem and of the Green-Kubo formula for the thermal conductivity. The generator of the dynamics is,
\qq
L=L_0+L_\tau
\label{generator}
\qqq
with,
\qq
L_0=\sum_i-\gamma p_i\frac{\partial}{\partial p_i}-\frac{\partial H}{\partial q_i}\frac{\partial}{\partial p_i}+\frac{\partial H}{\partial p_i}\frac{\partial}{\partial q_i}+\gamma T\frac{\partial^2}{\partial p^2_i}
\qqq
and
\qq
L_\tau=\frac{\tau}{2T}\sum_{i,k}(F^k(q_{i-k}-q_{i})+F^k(q_i-q_{i+k}))\frac{\partial}{\partial p_i}
\qqq
We note the basic identity,
\qq
-L^T_\tau H=L_\tau H=\frac{MJ\tau}{T}
\label{basic}
\qqq
where $L_\tau^T$ is the adjoint of the operator $L_\tau$ with respect to the Lebesgue measure.
We assume the existence, uniqueness, smoothness and regularity in the parameter $\tau$ of a stationary measure satisfying, $L^T\rho^\tau=0$.
\vskip 3mm
\noindent {\bf Remark.} The structure of the additional force proportional to the parameter $\tau$ is chosen so that (\ref{basic}) is satisfied, or more generally so that the generalized detailed balance relation  holds, see (\ref{detailed}) below. This in turn implies the validity of the time-reversal characterization of the non-equilibrium dynamics (\ref{reversalbasic}) (and (\ref{reversalbasic1}), see below).  The only degree of freedom left is the value of the friction parameter $\gamma$ which does not enter in (\ref{reversalbasic}) and (\ref{reversalbasic1}) as it does not  in the stationary measure at equilibrium either.  It should be seen as a regularizing parameter.  Indeed as we shall see explicitely in the last section, its main role is to provide an ``external" degradation mechanism for the energy current in the chain.  The physical degradation should come however from the nonlinearity of the chain and we shall take the friction $\gamma\rightarrow 0$ in order to evaluate the conduction properties of the chain. But {\it before} taking the limit $\gamma\rightarrow 0$, we will always take the limit $M\rightarrow\infty$.  By doing this, we will be able to discriminate between normal heat conduction and anomalous conduction, in one case the conductivity is infinite in the other it is not.  It is in the limit of $\gamma\rightarrow 0$ that we should expect the system defined by the dynamics (\ref{dynamics}) and the local dynamics of the original non-equilibrium problem to be equivalent.

\noindent We first study the evolution of the Shannon entropy under the evolution defined by the generator $L$.
To any measure with a smooth density $f$ on the phase space ${\bf R}^{2M}$ we associate the Shannon entropy,
\qq
S(f)=-\int dx f\log f
\qqq
with the notation $x=(\un p,\un q)$.
Compute its derivative with respect to time for a density whose evolution is given by
\qq
\partial_t f_t=L^Tf_t
\qqq
Next,
\beqnarray
\partial_tS(f_t)&=&-\int dx(1+\log f_t) (\partial_t f_t)\\
&=&-\int dx (1+\log f_t)L^Tf_t\\
&=&-\int dx f_t L\log f_t
\eeqnarray
\beqnarray
\partial_tS(f_t)&=&-\int dx f_t L(\log \frac{f_t}{\rho^0})+\beta\int dx f_t LH
\label{enteq}
\eeqnarray
where $\rho^0=Z^{-1}\exp(-\beta H)$.
For the first term, we note the identity,
\qq
L(\log h)=h^{-1}Lh-h^{-2}\Gamma(h,h)
\qqq
where $\Gamma$ is the Dirichlet form associated to the generator $L$,
\qq
\Gamma(h,g)=\sum_i \gamma T\frac{\partial h}{\partial p_i}\frac{\partial g}{\partial p_i}
\qqq
So that (\ref{enteq}) becomes, with the notation $g_t=f_t(\rho^0)^{-1}$
\qq
\partial_tS(f_t)=-\int dx \rho^0 L((\rho^0)^{-1} f_t)+\int dx \rho^0g_t^{-1}\Gamma(g_t,g_t)+\beta\int dx f_t LH
\label{enteq2}
\qqq
Integration by parts of the first term yields,
\qq
\partial_tS(f_t)=-\int dx ((\rho^0)^{-1} f_t)L^T\rho^0+\int dx \rho^0g_t^{-1}\Gamma(g_t,g_t)+\beta\int dx f_t LH
\qqq
Using (\ref{basic}), one gets $L^T\rho^0=\rho^0\frac{MJ\tau}{T^2}$ and therefore,
\qq
\partial_tS(f_t)=-\int dx  f_t\frac{MJ\tau}{T^2}+\int dx \rho^0g_t^{-1}\Gamma(g_t,g_t)+\beta\int dx f_t LH
\qqq
When $f_t$ is the density of the stationary measure $\rho^\tau$, the last term drops and one gets,
\qq
0=-\int dx  \rho^\tau(x)\frac{MJ\tau}{T^2}+\int dx \rho^0(x) \Gamma((\rho^0)^{-1}\rho^\tau,(\rho^0)^{-1}\rho^\tau)\frac{\rho^0}{\rho^\tau}
\label{lastent}
\qqq
 It is natural to define the  entropy production as
\qq
\sigma\equiv\frac{MJ\tau}{T^2}
\qqq
And using the fact that the second term of (\ref{lastent}) is given by a (positive) Dirichlet form which vanishes iff $\rho^\tau=\rho^0$, one concludes that the average entropy production in the stationary state is positive and that the current $J\neq 0$ iff $\tau\neq 0$.
\section{Generalized detailed balance relation.}
\subsection{Fluctuation theorem for the entropy production.}

Consider the ergodic average of the entropy production,
\qq
\bar\sigma_t\equiv\frac{1}{t}W(t)\equiv\frac{1}{t}\int_0^t\sigma(s)ds=\frac{1}{t}\int_0^t\frac{M\tau}{T^2}J(\un q(s),\un p(s))ds.
\qqq
If the process is ergodic one has,
\qq
\lim_{t\rightarrow\infty}\bar\sigma_t=\left<\sigma\right>_{\rho^\tau}
\qqq
for $\rho^\tau$-almost every initial conditions $\un q(0),\un p(0)$.
The fluctuation theorem is concerned with the specific shape of the large deviation functional $ e(w)$ of $\bar\sigma_t $ defined roughly as,
\qq
{\bf P}(\bar \sigma_t \in [w-\epsilon,w+\epsilon])\sim\exp(-e(w)t)
\qqq
The Gallavotti-Cohen fluctuation theorem states that
\qq
e(w)-e(-w)=-w.
\qqq
and by standard arguments \cite{Eckent,Kurchan,LebowitzSpohn,ReybelletThomas1}, assuming good properties of the process, the validity of this identity is basically ensured by the {\it generalized detailed balance relation} which is elementary to check in our case. Indeed, defining the reversal of velocities operator $\pi$, $\pi f(\un p,\un q)=f(-\un p,\un q)$, one checks, for any $\alpha$,
\qq
(\rho^0)^{-1}\pi (L^T-\alpha\sigma)\pi\rho^0=L-(1-\alpha)\sigma.
\label{detailed}
\qqq

\subsection{Time-reversal and entropy production.}

We now show  that the dynamics (\ref{dynamics}) has the property (\ref{reversalbasic}), expressing the fact the time-reversal symmetry is broken in a proper way.
In order to compare the space-time probability of a process and of its time-reverse, we first choose a reference process, to which we shall compare both.
Let $P_{\mu}^t$ the pathspace measure on trajectories $\xi=((\un q(s),\un p(s)), s\in [-t,t])$ obtained from the dynamics (\ref{dynamics}) started from initial conditions which are distributed according to some measure $\mu$.  We define the time-reversal operator on trajectories as follows, $(\Pi\xi)_t=\pi\xi_{-t}$, where $\pi$ as above reverse the sign of the momenta.  We want to evaluate the ratio $dP_{\mu}^t/dP_{\mu}^t\Pi$, where, as in section 2, $P_{\mu}^t\Pi$ is the distribution of the time-reverse process.
We compute first  $dP_{\mu}^t/dP_{\rho^0}^t $, where $P_{\rho^0}^t$ is the pathspace measure obtained from the dynamics (\ref{dynamics}) with $\tau=0$ and as initial distribution the Gibbs measure $\rho^0=Z^{-1}\exp(-\beta H)$.
A direct application of Girsanov formula yields,
\qq
dP_{\mu}^t=\exp(A^t(\xi))dP_{\rho^0}^t
\label{girsanov1}
\qqq
where,
\beqnarray
A^t(\xi)&=&\int_{-t}^t\frac{\tau}{4\gamma T^2}\un F(\un q(s)) d\un p(s)-\int_{-t}^t\frac{\tau^2}{16\gamma T^3}\un F^2(\un q(s)) ds
\nonumber
\\
&+&\int_{-t}^t\frac{\tau}{4 T^2}\un F(\un q(s)) \un p(s) ds+ \ln \mu(\xi_{-t})-\ln\rho^0(\xi_{-t}),
\label{girsanov2}
\eeqnarray
with, $F_i(\un q)=\sum_k(F^k(q_i-q_{i-k})-F^k(q_i-q_{i+k}))$ and the first term of (\ref{girsanov2}) is an Ito integral. Using the reversibility of the equilibrium measure, i.e. $P_{\rho^0}^t=P_{\rho^0}^t\Pi$, we write,
\qq
\frac{dP_{\mu}^t}{dP_{\mu}^t\Pi}=\frac{dP_{\mu}^t}{dP_{\rho^0}^t}\frac{dP_{\rho^0}^t}{dP_{\mu}^t\Pi}=\exp(A^t(\xi)-A^t(\Pi\xi))
\label{action}
\qqq
Being invariant under the time reversal operator $\Pi$, the second term in (\ref{girsanov2}) does not contribute to the difference in the above (\ref{action}).  The fact that the first term does not contribute is somewhat less obvious, see for instance example 3 in \cite{Maes1} for a proof of this fact in a similar set-up.
Finally, using $\un F(\un q(s)) \un p(s)=2MJ(s)$, we get,
\qq
R^t(\xi)\equiv A^t(\xi)-A^t(\Pi\xi)=\int_{-t}^t\sigma(s) ds +\beta H(\xi_{-t})-\beta H(\xi_t),
\label{reversalbasic1}
\qqq
modulo the boundary terms involving the distribution $\mu$, this is the same as (\ref{reversalbasic}) with the identification $\tau=N^{-1}\nabla T(x)$, $\beta=T^{-1}(x)$ and $M=[\epsilon N]$.
\section{Properties of the stationary state.}
\subsection{Heat conductivity and Green-Kubo formula.}
We define the heat conductivity $\kappa$ as
\qq
\kappa\equiv\lim_{\gamma\rightarrow 0}\lim_{M\rightarrow\infty}\lim_{\tau\rightarrow 0}\tau^{-1}\left<J\right>_{\rho^\tau}
\label{defconductivity}
\qqq 
and show now that it formally coincides with the usual Green-Kubo formula for Hamiltonian systems.
The Green-Kubo formula for the thermal conductivity of an Hamiltonian system is,
\qq
\kappa_{GK}=\lim_{M\rightarrow\infty}\frac{M}{T^2}\int_0^\infty\big<J(0)J(s)\big>_{\rho^0} ds.
\label{greenk}
\qqq
where $J(s)$ is given by (\ref{spatial}) in terms of the coordinates $\un q(s), \un p(s)$, solution of the Hamiltonian {\it deterministic} equations (i.e (\ref{dynamics}) with $\tau=0$ and $\gamma=0$) with initial conditions $\un q(0), \un p(0)$. 

\noindent As before, we assume the existence and uniqueness of a smooth (both in space and as function of $\tau$) invariant measure for the process (\ref{dynamics}), whose density $\rho^{\tau}$  satisfies,
\qq
L^T\rho^{\tau}=0
\label{sta}
\qqq
Developping at first order in $\tau$ the density $\rho^{\tau}=\rho^{0}+\tau\rho^1+\ldots$, we get an equation for the first-order correction to equilibrium (at temperature $T$) from,
\qq
L^T\rho^{\tau}=L^T_{0}\rho^0+\tau L^T_{0}\rho^1+L^T_{\tau}\rho^{0}+\ldots
\qqq
and thus, formally,
\qq
\rho^1=-(L^T_{0})^{-1}L^T_{\tau}\rho^{0}
\label{correction}
\qqq

\noindent We first compute $\lim_{\tau\rightarrow 0}\tau^{-1}\left<J\right>^{\tau}$.  Using (\ref{correction}) and again (\ref{basic})
\qq
\rho^1=-(L^T_{0})^{-1}\frac{M}{T^2}J\rho^{0}.
\qqq
Next, because the expected value of the current is zero in equilibrium and $((L^T_{0})^{-1})^T=L^{-1}_{0}$, we get,
\qq
\lim_{\tau\rightarrow 0}\frac{\big<J\big>^{\tau}}{\tau}=-\frac{M}{T^2}\big<J(L_{0})^{-1}J\big>_{\rho^0}=\frac{N}{T^2}\int_0^\infty\big<J(0)\big<J(s)\big>^{\gamma,T}\big>_{\rho^0} ds
\qqq
where $\big<J(s)\big>^{\gamma,T}$ is the solution of 
\qq
\frac{d}{ds}\big<J(s)\big>^{\gamma,T}=L_{0}\big<J(s)\big>^{\gamma,T}
\qqq
with initial conditions $\big<J(0)\big>^{\gamma,T}=J(0)$.
Formally, the most natural thing to do is to take next the limit of  $M\rightarrow\infty$ and then the limit $\gamma\rightarrow 0$ to recover the Green-Kubo formula (\ref{greenk}).  So that we get finally our definition of the conductivity (\ref{defconductivity}).
\vskip 3mm

\noindent{\bf Remark.} The discussion of these limits is a bit delicate in the case of {\it anomalous} conductivity in which case one tries to evaluate the dependance in $M$ of the conductivity $\kappa$, proceeding as we do yield $\kappa=\infty$ each time that there is anomalous conductivity. If one wishes to study the size dependance of the conductivity as for instance in \cite{Bernardin1,Pereverzev}, it seems that one should take $\gamma$ scaling as $M^{-1}$ and let go $M\rightarrow\infty$. But at the moment, we do not have any justification for that particular scaling of the friction and suspect that it might be model-dependent while the definition (\ref{defconductivity}) is not and does discriminate between normal and anomalous heat conduction.

\subsection{Temperature profile.}
\noindent We come back now to the original problem of the chain heated at its boundaries.  As we have seen above, if the conductivity is finite, one may write,
\qq
\left<J\right>^\tau=\kappa(T) \tau
\label{fourier2}
\qqq
where $\kappa$ will in general depend not only on the temperature $T$ but also on the physical parameters of interaction, in the limit of large $N$ and small $\gamma$.  As we shall illustrate in the next section on a concrete example, one obtains such a relation by simply writing down the stationary Fokker-Planck equation for the dynamics (\ref{dynamics}) and analyzing the effect of the nonlinearity on the nonequilibrium properties.  But now, as we explained at the end of section 2, we do the identification $\tau=N^{-1}\nabla T(x)$, $T=T(x)$, $M=[\epsilon N]$ so that (\ref{reversalbasic1}) is really identical to (\ref{reversalbasic}) . And we get,
\qq
\left<j_i\right>=\left<J\right>^\tau=\kappa(T(x))\frac{1}{N}\nabla T(x).
\qqq
for $i\in B_\epsilon(x)$, the box of size $\epsilon$ around the point $x$.  This is where one explicitely uses the assumption of equivalence between the original local dynamics (in the box $B_\epsilon(x)$) and of the dynamics (\ref{dynamics}) in the limit $\gamma\rightarrow 0$ because of their identical time-reversal property (\ref{reversalbasic}), (\ref{reversalbasic1}).
As we observed in (\ref{conservation000}), in the stationary state, the energy current must be constant along the chain and therefore, one gets,
\qq
\nabla\kappa(T(x))\nabla T(x)=0,
\label{profile}
\qqq
away from the boundaries.  The temperature profile is finally fixed by imposing the boundary conditions, which also sets the value of the current as a function of the imposed temperature difference.
\section{Models for heat conduction and current degradation.}
Having investigated the general properties of the stationary measure of the dynamics (\ref{dynamics}) and  shown that  its spacetime extension  has the natural characterization (\ref{reversalbasic}),(\ref{reversalbasic1}), we now study the stationary measure of a concrete model by writing down the equations that the correlation functions satisfy.  As a generic model for anharmonic systems having a normal conductivity, we consider a  lattice system described by the Hamiltonian,
\qq
H(\un q,\un p)=\sum_{i=1}^M \big[\frac{p_i^2}{2}+\omega^2\mu^2\frac{q_i^2}{2}+\frac{\lambda}{4}q_i^4+\frac{\omega^2}{2}(q_i-q_{i-1})^2\big]
\label{Hamilton}
\qqq
and the non-equilibrium dynamics (\ref{dynamics}) with periodic b.c. becomes,
\beqnarray
dq_i &=& p_i dt
\nonumber
\\
dp_i&=&-\gamma p_i dt -\omega^2((2+\mu^2)q_i-q_{i-1}-q_{i+1})-\lambda q_i^3\nonumber
\\
&+&\frac{\omega^2\tau}{2T}(q_{i+1}-q_{i-1})+\sqrt{2\gamma T}dw_i
\label{hardyn}
\eeqnarray

\noindent As above, one defines a local energy function,
\qq
h_i(\un p,\un q)=\frac{p_i^2}{2}+\omega^2\mu^2\frac{q_i^2}{2}+\frac{\lambda}{4}q_i^4+\frac{\omega^2}{4}(q_i-q_{i-1})^2+(q_i-q_{i+1})^2
\qqq
and the heat current in the chain $j_i$,
\qq
j_i=\frac{\omega^2}{2}(q_{i+1}-q_i)(p_i+p_{i+1})
\label{currentx}.
\qqq
and because of translation invariance, we shall often use the notation,
\qq
j\equiv\left<j_i\right>^\tau.
\qqq
for the average microscopic current in the stationary state.
\noindent Compute now the evolution of the current $j_i$ under the Hamiltonian evolution,
\qq
\frac{d^h j_i}{dt}=\frac{\omega^2}{2}((p^2_{i+1}-p^2_i)-(q_{i+1}-q_i)\omega^2((2+\mu)(q_i+q_{i+1})-(q_{i+1}+q_{i+2})-(q_{i-1}+q_i))+\lambda(q_i^3+q_{i+1}^3))
\qqq
As the dynamics is translation invariant, we can assume that the unique stationary measure is also translation invariant, which implies that,
\qq
\left<\frac{d^h j_i}{dt}\right>^\tau=\lambda\left(\left<q_iq_{i+1}^3\right>^\tau-\left<q^3_iq_{i+1}\right>^\tau\right)
\qqq
One obtains in the stationary state,
\qq
\frac{d}{dt}\left<j_i\right>=\left<\frac{d^h j_i}{dt}\right>^{\tau}-\gamma\left<j_i\right>^{\tau}+\frac{\omega^4\tau}{4T}\left<(q_{i+1}-q_{i-1})^2\right>^\tau=0
\qqq

So that we finally get,
\qq
-\gamma\left<j_i\right>^{\tau}+\frac{\omega^4\tau}{4T}\left<(q_{i+1}-q_{i-1})^2\right>^\tau+\lambda\left(\left<q_iq_{i+1}^3\right>^\tau-\left<q^3_iq_{i+1}\right>^\tau\right)=0
\label{basiccorrelations}
\qqq
This is the basic equation for the current.  The first term is a damping term for the current, the second term is a source term for the current (as we see below) and has the same sign as the parameter $\tau$, it comes directly from the irreversible part of the non-equilibrium dynamics, i.e it is produced by the force proportionnal to $\tau$ in (\ref{dynamics}).  The last term comes from the Hamiltonian part of the dynamics.  It should be responsible for the ``internal" degradation of the current in case of a normal conductivity.
\vskip 3mm
\noindent{\it Harmonic case.}

\noindent We look first at the harmonic case ($\lambda=0$).  By a direct computation (i.e. by checking $L^T\rho=0$ where $L^T$ is the adjoint of the operator defined in (\ref{generator})), one can show that the measure over the phase space ${\bf R}^{2M}$ with Gaussian density,
\qq
\rho(\un q,\un p)=Z^{-1}\exp\left(-\beta H(\un q, \un p)+\frac{M\tau}{\gamma T^2}J(\un q, \un p)\right)
\label{harmeasure}
\qqq
is stationary for the process defined by (\ref{hardyn}) when $\lambda=0$.
We now look directly at the equation (\ref{basiccorrelations}) with $\lambda=0$ and at the lowest-order in $\tau$ since we are interested in computing the heat conductivity, i.e we write,
\qq
-\gamma\left<j_i\right>^{\tau}+\frac{\omega^4\tau}{4T}\left<(q_{i+1}-q_{i-1})^2\right>^0=0
\label{hcorrelations}
\qqq
where $\left<.\right>^0$ denotes the expectation value with respect to the equilibrium harmonic measure.  This last term is readily computed and (\ref{hcorrelations}) yield,
\qq
\left<j_i\right>^{\tau}=\frac{\omega^2\tau}{\gamma}\int_{-\frac{1}{2}}^{\frac{1}{2}}\frac{\sin^2(2\pi x)}{\mu^2+4\sin^2(\pi x)}dx
\label{solharm}
\qqq
for large $N$.
Here the basic problem of harmonic chains is apparent, when the external friction $\gamma$ is taken to zero  the current becomes infinite, the harmonic chain has no internal mechanism which could degrade the current that is continuously created by the ``thermal force" proportional to $\tau$ (i.e by the local temperature gradient).
According to our definition, the conductivity of the harmonic model is,
\qq
\kappa=\lim_{\gamma\rightarrow 0}\lim_{M\rightarrow\infty}\lim_{\tau\rightarrow 0}\tau^{-1}\left< J\right>^\tau=+\infty
\qqq
Let us see now how to deal with the anharmonic interactions. Observe first that if the stationary measure was Gaussian, the conductivity of the system would be infinite, indeed under Gaussian assumption and translation invariance,
\qq
\left(\left<q_iq_{i+1}^3\right>^\tau-\left<q^3_iq_{i+1}\right>^\tau\right)=3(\left<q_iq_{i+1}\right>^\tau\left<q_i^2\right>^\tau-\left<q_iq_{i+1}\right>^\tau\left<q_{i+1}^2\right>^\tau)=0
\qqq
and one gets (\ref{hcorrelations}) again.
Observe also that $\left<q_iq_{i+1}^3\right>^\tau-\left<q^3_iq_{i+1}\right>^\tau$ is odd under the exchange of the indices $i$ and $i+1$, as is the current, by definition (\ref{currentx}).  
\vskip 3mm
\noindent{\it Stochastic approximation of the non-linearity.}

\noindent The simplest possible assumption which can be made about the stationary state is to assume,
\qq
\left<q_iq_{i+1}^3\right>^\tau-\left<q^3_iq_{i+1}\right>^\tau=-\eta j_i
\qqq
for some $\eta$.
The important point is that such a term may be produced by a dynamics which amounts to randomly exchange energy between nearest-neighbours in the chain and not by an external damping term in the dynamics.  This random exchange is responsible for the degradation of the current constantly produced by the ``thermal force" proportional to $\tau$.  See \cite{Bernardin} for details and rigorous results on the model. Now, (\ref{basiccorrelations}) become
\qq
-\gamma\left<j_i\right>^{\tau}-\eta\left<j_i\right>^{\tau}+\frac{\omega^4\tau}{4T}\left<(q_{i+1}-q_{i-1})^2\right>^0=0
\label{scorrelations}
\qqq
and the conductivity is,
\qq
\kappa=\frac{\omega^2}{\eta}\int_{-\frac{1}{2}}^{\frac{1}{2}}\frac{\sin^2(2\pi x)}{\mu^2+4\sin^2(\pi x)}dx,
\qqq
when $\mu=0$, one gets $\kappa=\frac{\omega^2}{2\eta}$ which is the number obtained in \cite{Bernardin}\footnote{There is a slight difference coming from the fact that in \cite{Bernardin}, the definition of the current contains a part coming from the random exchange itself, for small $\eta$ this becomes negligible in the computation of the conductivity.}.  As the conductivity does not depend on the temperature $T$, from (\ref{profile}), one gets a linear profile. More sophisticated models and stochastic approximations have been devised to tackle the problem of anomalous heat conduction \cite{Bernardin1}.
\vskip 3mm
\noindent{\it Closure approximation.}

\noindent We refer to \cite{LefevereSchenkel} for details about the closure approximation on the model we are considering in this section, but we give the outline of the argument.  Observe that (\ref{basiccorrelations}) involves a four-point correlation function.  In order to interpret it as term providing a internal mechanism for the degradation of the current, one would like to show that it is simply proportional to the average current, namely a two-point correlation function.  To achieve this, the simplest thing to do is to assume that the stationary measure is Gaussian.  As we observed above, this obviously factorizes the four-point correlation in terms of the two-point correlation but does not produce any significant correction to (\ref{basiccorrelations}) with respect to the harmonic case.  The next simplest thing to do is to write down the hierarchy of equations satisfied by the $n$-point correlation functions and assume that the six-point correlations factorize in terms of the two-point correlations.  After linearizing around the equilibrium harmonic solution, this yields a linear expression of the four-point correlations in terms of the two-point correlations.   The main point of the analysis is to identify the resonances in the interactions between the harmonic degrees of freedom (the phonons).  To formulate the end result of the closure approximation we need to consider the general set of correlation functions,
\qq
\hat J_l\equiv\omega^2\left<p_iq_{i+l}\right>^\tau
\qqq
the fact the RHS does not depend on $i$ follows from translation-invariance.  In particular, $\hat J_1=j$. We denote by $\un{\hat J}$ the vector made of the $\hat J_l$'s.
Now, after closure, the stationary equation for this vector takes the form, for large $\mu$,
\qq
-\gamma\un{\hat J}-\frac{\lambda^2T^2}{\omega^7\mu^5}{\cal L}(\un{\hat J})+\un\sigma=0
\qqq
Where ${\cal L}$ is a linear operator which does not depend on any physical parameter entering the equations of motion.  The components of the vector $\un \sigma$ coming from the thermal force are given by,
\qq
\sigma_m=\omega^2\tau\int_{-\frac{1}{2}}^{\frac{1}{2}}\frac{\sin(2\pi x)\sin(2\pi m x)}{\mu^2+4\sin^2(\pi x)}dx.
\qqq
When computing $\un \sigma$ we have neglected correction of order $\lambda$ in the equilibrium measure has been neglected.
When $\gamma\rightarrow 0$,  one inverts ${\cal L}$ and gets,
\qq
\un{\hat J}=\frac{\omega^7\mu^5}{\lambda^2T^2}{\cal L}^{-1}(\un \sigma)
\qqq
from which one may obtain $\hat J_1=j$ and thus the conductivity which is (for large $\mu$)
\qq
\kappa=\alpha\frac{\omega^9\mu^3}{\lambda^2T^2}
\qqq
The value of $\alpha=0.275637$ was computed in \cite{AokiSpohn} where the expression  (\ref{conductivity}) was also tested and numerically confirmed.  As for the temperature profile, one gets,
\qq
\nabla\frac{1}{T^2(x)}\nabla T(x)=0
\label{profilea}
\qqq
away from the boundaries.  The solution of (\ref{profilea}) and the  problem of its boundary conditions has been discussed in \cite{AokiSpohn}.

\vskip 3mm
Finally, we note once again that our method is not limited to the example we are treating here.  Each model will have a particular source term in the equation governing the degradation of the current.  This source term can be directly computed from the dynamics (\ref{dynamics}).  In particular, it would be very interesting to study the heat conductivity of the Toda lattice with this method, which we plan to do in a future publication.

\noindent {\bf Acknowledgments.}  It is a pleasure to thank C\'edric Bernardin, Thierry Bodineau, Alain Schenkel and Herbert Spohn for useful discussions.

\section{Appendix: Solution of the Harmonic case.}
We take the process defined by (\ref{hardyn}) with $\lambda=0$ and solve the equations for the correlations in the stationary state.

\noindent It is most convenient to work in coordinates where the equations become diagonal, and we introduce the  Fourier coordinates for the periodic harmonic chain by 
\qq
Q_k=\frac{1}{\sqrt{N}}\sum_{j=1}^{N} e^{i \frac{2\pi}{N}k j}q_j
\qqq
with $-N/2+1\leq k\leq N/2$. The $P_k$ coordinates are defined in a similar fashion.  We note for further purposes that $\ov Q_k=Q_{-k}$ and $\ov P_k=P_{-k}$.
In those coordinates, the Hamiltonian reads,
\qq
H(\un Q,\un P)=\sum_{k=-N/2+1}^{N/2} \big[\frac{|P_k^2|}{2}+\omega^2_k\frac{|Q_k|^2}{2}\big]
\label{HamiltonF}
\qqq
where 
\qq
\omega^2_k=\omega^2(\mu^2+4\sin^2(\frac{\pi k}{N}))
\qqq
The spatial average of the energy current in the periodic chain (\ref{spatial}) reads now
\qq
J=\frac{1}{N}\sum_{k =-N/2+1}^{N/2}\sin(\frac{2\pi k}{N}) \,{\rm Im}\left(Q_{-k} P_{k}\right).
\label{current}
\qqq
In Fourier coordinates, the equations (\ref{hardyn}) with $\lambda=0$ become,
\beqnarray
dQ_{k}&=&P_k dt
\label{HS1}
\\
dP_{k}&=&-\gamma P_k dt-(\omega^2_k-i\frac{\tau}{T}\sin(\frac{2\pi k}{N}))Q_k dt+\sqrt{2\gamma T}dW_k
\label{HS2}
\eeqnarray
The $W_k$ are complex Wiener processes which satisfy $\ov W_k=W_{-k}$ and, $\left<W_k,W_l\right>=\delta(l+k)$.

The equations for the correlations in the stationnary state are,
\beqnarray
\left<P_kQ_{-k}\right>+\left<Q_kP_{-k}\right>&=&0
\\
\left<|P_k|^2\right>-\omega^2_k\left<|Q_k|^2\right>&=&0
\\
2\gamma\left<|P_k|^2\right>+i\omega^2_k\alpha_k\frac{\tau}{T}\left<P_kQ_{-k}\right>&=&2\gamma T
\\
2\gamma\left<P_kQ_{-k}\right>-i\omega^2_k\alpha_k\frac{\tau}{T}\left<|Q_k|^2\right>&=&0
\eeqnarray
where $\alpha_k=2\omega^2\omega_k^{-2}\sin(\frac{2\pi k}{N})$.
These are readily solved and yield
\beqnarray
\left<|P_k|^2\right>&=&\delta_k T
\\
\left<P_kQ_{-k}\right>&=& \hf \gamma^{-1}i \delta_k \alpha_k\tau
\eeqnarray
with, $\delta_k=(1-\frac{\omega_k^2\alpha_k^2\tau^2}{4\gamma^2T^2})^{-1}$.
Keeping the lowest-order in $\tau$ and doing a little computation gives (\ref{solharm}) again.


\begin{thebibliography}{10}

\bibitem{AokiSpohn}K.~Aoki, J.~Lukkarinen, H.~Spohn: {\it Energy Transport in Weakly Anharmonic Chains.} Preprint arXiv.org:cond-mat/0602082, to appear in J.Stat. Phys.

\bibitem{Bernardin}C.~Bernardin, S.~Olla:{\it Fourier's law for a microscopic model of heat conduction.} J. Stat. Phys. {\bf 121},827-852 (2005)

\bibitem{Bernardin1}G. Basile, C. Bernardin, S. Olla :{\it Momentum conserving model with anomalous thermal conductivity in low dimensional systems.}  Phys.Rev.Lett. {\bf 96}, 204303 (2006)

\bibitem{Bon}F.~Bonetto, J.~L.~Lebowitz, J.~Lukkarinen: {\it Fourier's law for a harmonic crystal with self-consistent stochastic reservoirs.}
J.~Stat.~Phys.~{\bf 116}, 783--813 (2004).

\bibitem{bonetto} F.~Bonetto, J.~L.~Lebowitz, L.~Rey-Bellet : {\it Fourier's law: a challenge to theorists.} {\it Mathematical Physics 2000}, 128--151, Imp.~Coll.~Press, 2000.

\bibitem{Bricmont}J.~Bricmont, A.~Kupiainen :{On the derivation of Fourier's law for coupled anharmonic oscillators.}http://arxiv.org/abs/math-ph/0605062

\bibitem{eck1}J.-P. Eckmann, C.-A. Pillet, L. Rey-Bellet:
     {\it Non-equilibrium statistical mechanics of anharmonic chains coupled to two heat baths at different temperatures.}
     Commun. Math. Phys. {\bf 201}, 657--697 (1999)


\bibitem{Eckent} J-P.~Eckmann, C-A.~Pillet, L.~Rey-Bellet:{\it Entropy production in Non-linear, Thermally Driven Hamiltonian Systems.}J. Stat. Phys. {\bf 95}, 305--331 (1999)

\bibitem{eck3} J.-P. Eckmann, M. Hairer:
{\it Non-Equilibrium Statistical Mechanics of Strongly Anharmonic Chains of Oscillators.}
Commun. Math. Phys. {\bf 212}, 105--164 (2000)


\bibitem{Evans} D.~J.~Evans, G.~P.~Morriss: {\it Statistical Mechanics of Nonequilibrium Liquids}, Academic Press, San Diego, 1990.

\bibitem{Horie}C. Horie, J. A. Krumhansl:{\it Boltzmann equation in a Phonon system.} Phys. Rev. {\bf 136}, A1397ÐA1407 (1964)

\bibitem{Kipnis}C. ~Kipnis, C.~ Marchioro, E.~Presutti :{\it Heat flow in an exactly solvable model.}
J. Statist. Phys. {\bf 27}  65--74 (1982).

\bibitem{Kurchan}J.~Kurchan: {\it Fluctuation Theorem for stochastic dynamics.} Journal of Physics  {\bf A31},3719 (1998)

\bibitem{Kwok} P.C.K. Kwok :{\it Green's function method in lattice dynamics}  Solid state Phys. {\bf 20},213-303 (1967)
\bibitem{LefevereSchenkel} R.~Lefevere, A.~Schenkel :{\it Normal heat conductivity in a strongly pinned chain of anharmonic oscillators.}   J. Stat. Mech. (2006) L02001

\bibitem{LebowitzSpohn} J.L.~Lebowitz, H.~Spohn: {\it A Gallavotti-Cohen-type symmetry in the large deviation functional for stochastic dynamics.} J.Stat.Phys. {\bf 95}, 333-365 (1999)

\bibitem{livi} S.~Lepri, R.~Livi, A.~Politi: 
{Thermal conduction in classical low-dimensional lattices.} 
 Phys.~Rep.~{\bf 377}, 1--80 (2003).


\bibitem{Maes0} C.~Maes :{\it Fluctuation theorem as a Gibbs property.} J.Stat.Phys. {\bf 95}, 367-392 (1999)

\bibitem{Maes1}C.~ Maes, F.~ Redig and A. ~Van Moffaert :{\it On the definition of entropy production, via examples} , J. Math. Phys.{\bf  41}, 1528-1554 (2000)

\bibitem{Maes2}C. ~Maes, K. ~Netocny  and M. ~Verschuere:   {\it Heat Conduction Networks.}   J. Stat. Phys. {\bf 111}, 1219-1244 (2003).


\bibitem{nak} H. Nakazawa:
{\it On the lattice thermal conduction.} Supp. Prog. Th. Physics {\bf 45},
231--262 (1970)

\bibitem{Peierls} R. Peierls:{\it Zur kinetischen Theorie der Warmeleitung in Kristallen.} Ann. Phys. (Ger.){\bf 3}, 1055-1101(1929)

\bibitem{Pereverzev} A. Pereverzev :{\it Fermi-Pasta-Ulam  $\beta$-lattice: Peierls equation and anomalous heat conductivity.} Phys. Rev. {\bf E 68}, 056124 (2003)



\bibitem{ReybelletThomas0} L. Rey-Bellet, L.E. Thomas:
{\it Exponential Convergence to Non-Equilibrium Stationary States in
  Classical Statistical Mechanics.} 
 Commun. Math. Phys. {\bf 225}, 305--329 (2002)

\bibitem{rll} Z. Rieder, J.L. Lebowitz, E. Lieb:
{\it Properties of a Harmonic Crystal in a Stationary Nonequilibrium
  State.} J.~Math.~Phys. {\bf 8}, 1073--1078 (1967)

\bibitem{ReybelletThomas1}L.~Rey-Bellet,L.E.~Thomas: {\it Fluctuations of the entropy production in anharmonic chains .} Ann. Henri Poinc.{\bf 3} 483--502 (2002) 

\bibitem{Spohn} H.Spohn:{\it The phonon Boltzmann equation, properties and link to weakly anharmonic lattice dynamics}, 2005 Preprint math-ph/0505025, to appear in J.Stat.Phys.
\end{thebibliography}
\end{document}